\documentclass[12pt]{spieman}  
\usepackage{amsmath,amsfonts,amssymb}
\usepackage{graphicx}
\usepackage{setspace}
\usepackage{tocloft}

\usepackage{siunitx}
\usepackage{booktabs}
\usepackage{lineno}
\graphicspath{{./Figure/eps/}}

\newcommand{\um}{\,\si{{\micro}m}}
\newcommand{\mm}{\,\si{mm}}
\newcommand{\cm}{\,\si{cm}}
\newcommand{\m}{\,\si{m}}
\newcommand{\km}{\,\si{km}}
\newcommand{\eV}{\,\si{eV}}
\newcommand{\keV}{\,\si{keV}}

\title{Multi-image x-ray interferometer module: II. Demonstration of high-resolution x-ray imaging with regular-interval coded apertures}

\author[a]{Kazunori Asakura}          
\author[a,b,c]{Kiyoshi Hayashida}   
\author[d]{Tomokage Yoneyama}    
\author[a,b]{Hirofumi Noda}            
\author[a]{Marina Yoshimoto}          
\author[a]{Tomohiro Hakamata}      
\author[a,b]{Hironori Matsumoto}    
\author[a]{Hiroshi Tsunemi}             

\affil[a]{Department of Earth and Space Science, Graduate School of  Science, Osaka University, 1-1 Machikaneyama, Toyonaka, Osaka 560-0043, Japan}
\affil[b]{Project Research Center for Fundamental Sciences, Graduate School of Science, Osaka University, 1-1 Machikaneyama, Toyonaka, Osaka 560-0043, Japan}
\affil[c]{Institute of Space and Astronautical Science, Japan Aerospace Exploration Agency, 3-1-1 Yoshino-dai, Chuo-ku, Sagamihara, Kanagawa 252-5210, Japan}
\affil[d]{Faculty of Science and Engineering, Chuo University, 1-13-27 Kasuga, Bunkyo, Tokyo 112-8551, Japan}

\cftpagenumbersoff{figure}
\cftpagenumbersoff{table} 
\begin{document} 
\maketitle

\begin{abstract}

We have been developing an x-ray imaging system, Multi-Image X-ray Interferometer Module (MIXIM), to achieve a high angular resolution with a compact system size.
MIXIM is comprised of a mask with equally-spaced apertures and an x-ray detector. 
The aperture size and mask-detector distance determine the system's angular resolution.
Although a smaller aperture gives a better resolution, the degree of improvement is limited by a diffraction effect.
MIXIM circumvents this problem by utilizing the Talbot effect.
Our experiment with the previous model equipped with a multi-pinhole mask obtained an angular resolution of $\ang{;;0.5}$ with a mask-detector distance of $92\cm$.
A major downside of the multi-pinhole mask is, however, that it has a very low opening fraction, which results in a very low effective area.
Here, we newly adopt to MIXIM a multiple coded aperture (MCA) mask, an array of coded aperture patterns.
Our proof-of-concept experiment demonstrates that the Talbot effect works even for the MCA mask with a high opening fraction of $\sim50\%$ at $12.4\keV$.
Consequently, the new MIXIM realizes about 25 times as large an effective area as that of the previous model,
while maintaining a high angular resolution of $\ang{;;0.2}$ and a compact size of $\sim1.5\m$.

\end{abstract}


{\noindent \footnotesize\textbf{*}Kazunori Asakura,  \linkable{asakura\_k@ess.sci.osaka-u.ac.jp} }

\begin{spacing}{2}   

\section{Introduction}
\label{sec:intro}  

The angular resolution of an imaging system is the key measure for x-ray astronomy 
to resolve the spatial structure of astrophysical x-ray sources and diagnose their physical properties.
Many imaging systems onboard recent major x-ray astronomical satellites, e.g., \textit{ASCA}\cite{Tanaka1994}, \textit{Chandra}\cite{Weisskopf2000}, 
\textit{XMM-Newton}\cite{Jansen2001}, \textit{Suzaku}\cite{Mitsuda2007}, \textit{Swift}\cite{Gehrels2004}, and \textit{Hitomi}\cite{Takahashi2016}, consist of Wolter type-I telescopes and x-ray CCDs.
Notably, the angular resolution of the High Resolution Mirror Assembly (HRMA) onboard \textit{Chandra} is unrivaled, $\ang{;;0.5}$\cite{Weisskopf2000}, 
with which unprecedented scientific outputs have been produced since its launch in 1999\cite{Tananbaum2014}.
The performance of the Wolter type-I optics has room for further improvement in principle because the performance is so far not limited by the diffraction limit
but by the manufacturing precision, specifically the degree of the surface smoothness and alignment accuracy of the mirror shells.
However, it is practically unfeasible with current technology to manufacture and align mirror shells with a higher precision than the HRMA.

Meanwhile, x-ray interferometry has been developed as a promising alternative approach to achieve high angular resolution\cite{Uttley2021},
or potentially a much higher resolution than conventional methods can achieve.
\textit{MAXIM}\cite{Cash2003} and \textit{MAXIM Pathfinder}\cite{Gendreau2003}, two pioneering x-ray interferometry projects,
have goals of angular resolutions of $0.1$ and $100$ micro-arcseconds, respectively.
A prototype of \textit{MAXIM} with a system size of $100\m$ has already demonstrated that 
it can form interference fringes at $1.25\keV$ with an angular resolution of $\ang{;;0.1}$\cite{Cash2000}.
A \textit{MAXIM}-type interferometer can in principle further improve its angular resolution with an increased separation of mirrors,
though it needs to be accompanied by an increase in the distance between mirrors and a detector to make fringes resolvable.
Even the comparatively moderate-capability \textit{MAXIM Pathfinder} has to maintain a large mirror-detector distance of $200\km$ 
to achieve its angular-resolution goal, and \textit{MAXIM} requires a much greater distance. 
High-precision formation flight of multiple satellites has been proposed to realize such a long distance.
Although it should be feasible in principle, an actual deployment in orbit is not going to happen anytime soon because a variety of technical challenges still remain.
Alternative methods, such as x-ray interferometers with a slatted mirror\cite{Willingale2004} or a beam splitter with multi-layers\cite{Kitamoto2011, Kitamoto2014},
have been also suggested to reduce the system size, but the detection of fringes with such systems in the x-ray band has not yet been reported.
It follows that there is currently no way to obtain an angular resolution higher than $\ang{;;0.5}$ in x-rays with the scale of a single satellite ($<10\m$).

Motivated by the lack of ultra-high-resolution astronomical x-ray imaging observatories, 
we have been developing a novel high-resolution x-ray imaging system, Multi-Image X-ray Interferometer Module (MIXIM)\cite{Hayashida2016}.
MIXIM is comprised of a mask with equally-spaced multiple apertures and an x-ray detector.
Its imaging principle is basically the same as that of a pinhole camera, and accordingly, 
its angular resolution is determined by an aperture size and a distance between the mask and detector.
Although the angular resolution of a simple pinhole camera is limited by a diffraction effect,
MIXIM circumvents the limitation by diffraction with the application of the Talbot effect\cite{Talbot1836} and is able to obtain a high-resolution image at a particular wavelength 
(for the detailed principle, see our previous paper\cite{Asakura2023a}; hereafter, referred to as ``Paper I'').
Ideally, MIXIM should achieve an angular resolution of $<\ang{;;0.5}$ with a system size of as small as $50\cm$ at $12.4\keV$.

We already conducted proof-of-concept experiments with the previous model of MIXIM with a multi-pinhole mask and a fine-pixel CMOS sensor in SPring-8, a synchrotron radiation facility in Japan, 
and achieved angular resolutions of $\ang{;;0.5}$ and $\ang{;;0.05}$ with mask-sensor distances of $92\cm$ and $866.5\cm$, respectively\cite{Asakura2023a}.
The experiments demonstrated that MIXIM can simultaneously realize a high angular resolution and compact physical size.
Thus, MIXIM has a great advantage with regard to feasibility compared with other proposed x-ray interferometers.
A critical downside of MIXIM is, however, that the multi-pinhole system adopted for MIXIM has a transmittance of only $1.3\%$ at $12.4\keV$ due to a very low opening fraction.
Hence, it is not still practical for observation of astrophysical x-ray sources since they usually have low photon fluxes.

To solve this crucial problem, we devised a new method, employing multiple coded-aperture masks instead of multiple pinholes in MIXIM,
in which configuration the transmittance is increased by a factor of 20 or more.
With multiple coded-aperture masks, the Talbot effect should work and result in self-images, 
providing that the distance between the masks and detector is well controlled, depending on the wavelength of incoming photons;
then the effect can be utilized to reconstruct the original image, as explained in section \ref{sec:MCA}.
We made three sets of multiple coded-aperture masks, installed them to MIXIM, conducted proof-of-concept experiments (section \ref{sec:experiment}),
and obtained positive results, demonstrating that the Talbot effect does work and that high-resolution imaging is achieved (section \ref{sec:results}).
This is a novel, yet realistic and promising idea for compact high-resolution x-ray imagers for astronomy, as discussed in section \ref{sec:discussion}.

\section{Multiple Coded Aperture Masks}
\label{sec:MCA}  

In the field of hard x-ray and gamma-ray astronomy, an array of pinholes (such as masks with randomly-distributed holes and coded aperture masks) is often adopted instead of a single pinhole mask
because it increases an opening fraction while it retains an angular resolution equivalent to a single pinhole mask (see, e.g., the review Ref. \citenum{Caroli1987}).
Indeed, coded apertures are equipped with recent x-ray and gamma-ray observatories, including \textit{INTEGRAL}\cite{Ubertini2003} and \textit{Swift}\cite{Barthelmy2005},
which realize angular resolutions of $\ang{;10;}$--$\ang{;20;}$, as expected from their aperture sizes and mask-detector distances
(\textit{INTEGRAL} and \textit{Swift} have mask element sizes of $11.2\mm \times 11.2\mm$ and $5\mm \times 5\mm$, respectively).
Notably, reducing the mask and detector element sizes would further improve these angular resolutions to some extent,
though the improvement is ultimately limited due to the diffraction limit, as with the case of a pinhole mask\cite{Asakura2023a}.

\begin{figure} [ht]
   \begin{center}
   \begin{tabular}{c} 
   \includegraphics[height=10cm]{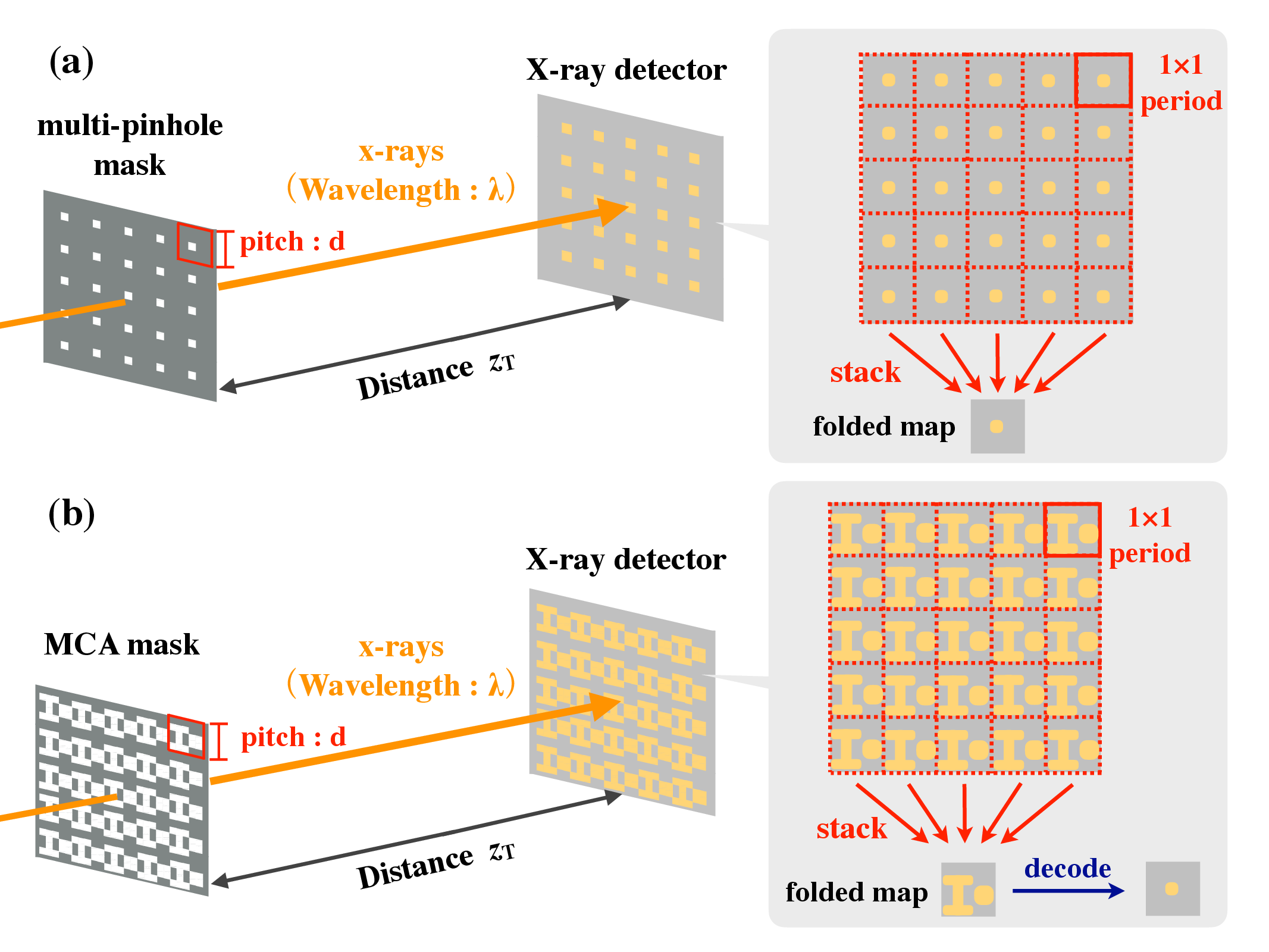}
   \end{tabular}
   \end{center}
   \caption[Schematic charts of the analysis procedures.]  
   { \label{fig:analysis_flow} 
   Schematic charts of (a) the analysis procedures with the previous model of MIXIM with a multi-pinhole mask 
   (see also Paper I), and (b) the newly proposed analysis procedures with the new MIXIM with a MCA mask.
   Whereas a MCA mask necessitates an additional decoding procedure, it substantially increases an opening fraction from the multi-pinhole mask configuration.}
   \end{figure}

In this work, we followed this approach and newly adopted an array of multiple coded apertures (MCAs) to MIXIM, 
replacing the currently-adopted multi-pinhole mask, to increase the opening fraction of the existing MIXIM system without sacrificing the high angular resolution.
Figure \ref{fig:analysis_flow} shows a schematic chart of the imaging procedures with MIXIM with a multi-pinhole mask and that with a MCA mask.
Equally-spaced aperture patterns are supposed to form self-images for monochromatic parallel light at the distance $z_\mathrm{T}$ according to the following formula\cite{Wen2013}:
\begin{linenomath}
\begin{align}
  \label{eq:distance_plane}
  z_{\mathrm{T}} = m\frac{d^2}{\lambda} \,(m = 1, 2, 3...),
\end{align}
\end{linenomath}
where $d$ is the arrangement pitch of the apertures and $\lambda$ is the wavelength of the incident light.
MIXIM folds the arranged self-images to obtain a single image with high photon statistics (see Paper I for the detailed procedures\cite{Asakura2023a}).
Notably, the Talbot effect ideally occurs with not only pinholes but arbitrary aperture patterns arranged at periodic intervals,
although the Talbot effect has never been experimentally demonstrated with complex aperture patterns at least in the x-ray band.
If the Talbot effect actually occurs with a MCA mask, the stacked image represents the convolution of an x-ray source profile and the unit aperture pattern.
This follows that the original source profile can be derived through a decoding procedure in a similar way to general coded-aperture imaging systems.

The angular resolution and field of view (FOV) of a MCA mask are approximated to be $rz^{-1}$ and $dz^{-1}$, respectively,
where $r$, $d$, and $z$ denote the mask element size, arrangement pitch of the apertures, and mask-detector distance, respectively.
In addition, the pitch determines the mask-detector distance for an application of the Talbot effect ($z_{\mathrm{T}} \propto d^2$),
which indicates that a fine-pitch MCA mask is required to realize simultaneously a high angular resolution and compactness.
In this work, we introduced three types of fine-pitch MCA masks, which were fabricated by the LIGA 
(the German acronym for lithography, electroplating, and molding) process at the Karlsruhe Institute of Technology.

\begin{figure} [ht]
   \begin{center}
   \begin{tabular}{c} 
   \includegraphics[height=7.5cm]{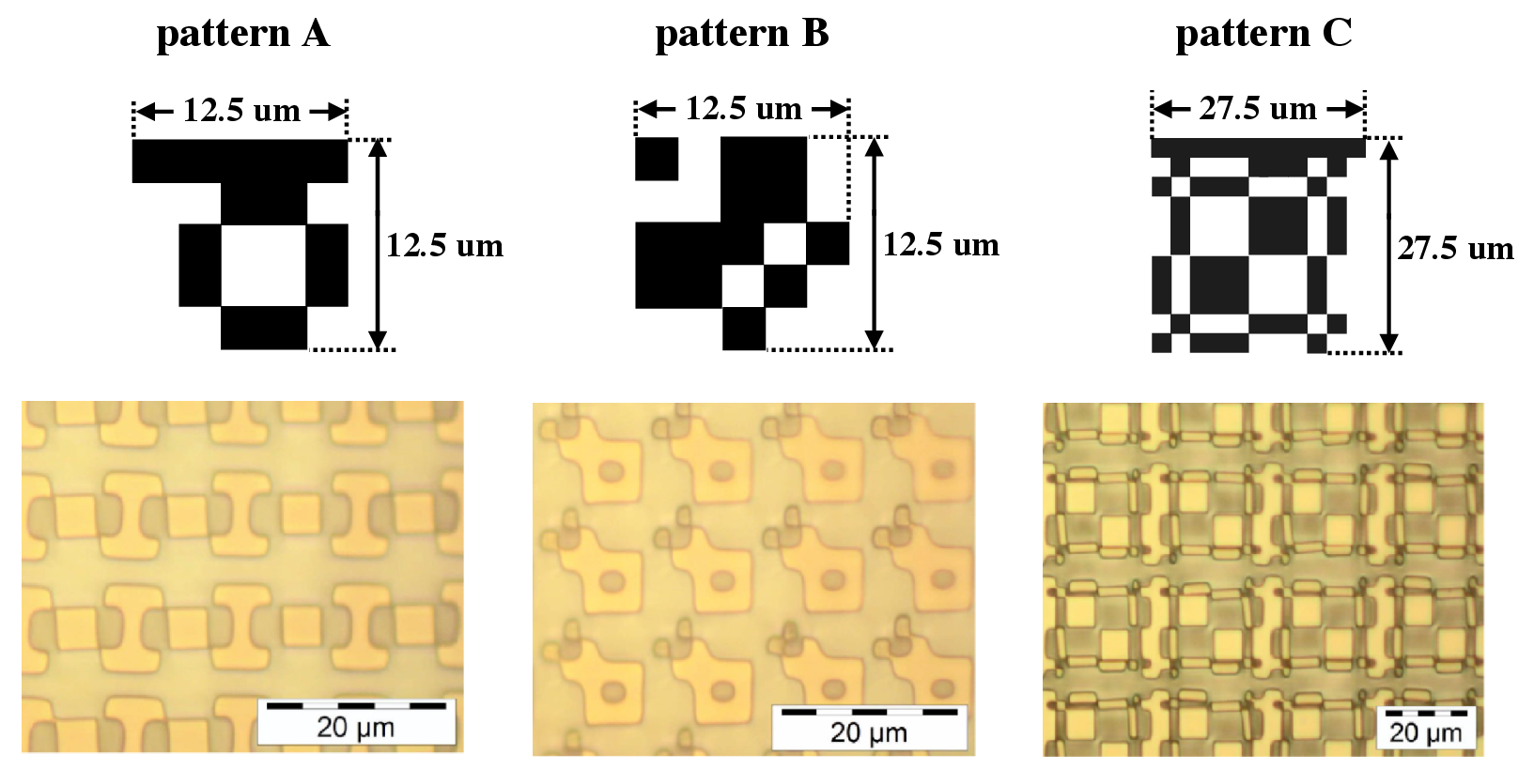}
   \end{tabular}
   \end{center}
   \caption[Original designs and actual micrographs of the MCA masks.]  
   { \label{fig:micrographs} 
   (Top) Original designs of the unit aperture patterns for the MCA masks, patterns A--C, 
   and (Bottom) actual micrographs of the MCA masks fabricated with the LIGA process.
   Gold absorbers are deposited on the substrates such that the unit aperture patterns are arranged at regular intervals.}
   \end{figure}

Figure \ref{fig:micrographs} illustrates the original designs and actual micrographs of the three MCA masks, designated as patterns A--C.
These masks consist of gold absorbers with a thickness of $>20\um$ and polyimide substrates (including chromium and gold) with a thickness of $550\um$.
Whereas all of them have the same size of $15\mm \times 15\mm$, their unit aperture patterns differ;
patterns A and B have a pitch of $12.5\um$ ($5\times5$ elements), while pattern C has a pitch of $27.5\um$ ($11\times11$ elements).
We note that patterns A and C are modified uniformly redundant arrays (MURAs), the optimal patterns designed by Gottesman and Fenimore\cite{Gottesman1989},
whereas pattern B is not (but its auto-correlation function also has a sharp peak, as with the MURAs).
The theoretical angular resolutions at $12.4\keV$ ($m=1$) are derived to be $\ang{;;0.3}$ and $\ang{;;0.07}$ for the pitches of $12.5\um$ and $27.5\um$, respectively;
thus, these MCA masks have a potential for high-resolution x-ray imaging.

\section{Proof-of-concept Experiments}
\label{sec:experiment}  

\subsection{Setups}
\label{sec:experiment_2}
\begin{figure} [ht]
   \begin{center}
   \begin{tabular}{c} 
   \includegraphics[height=5.7cm]{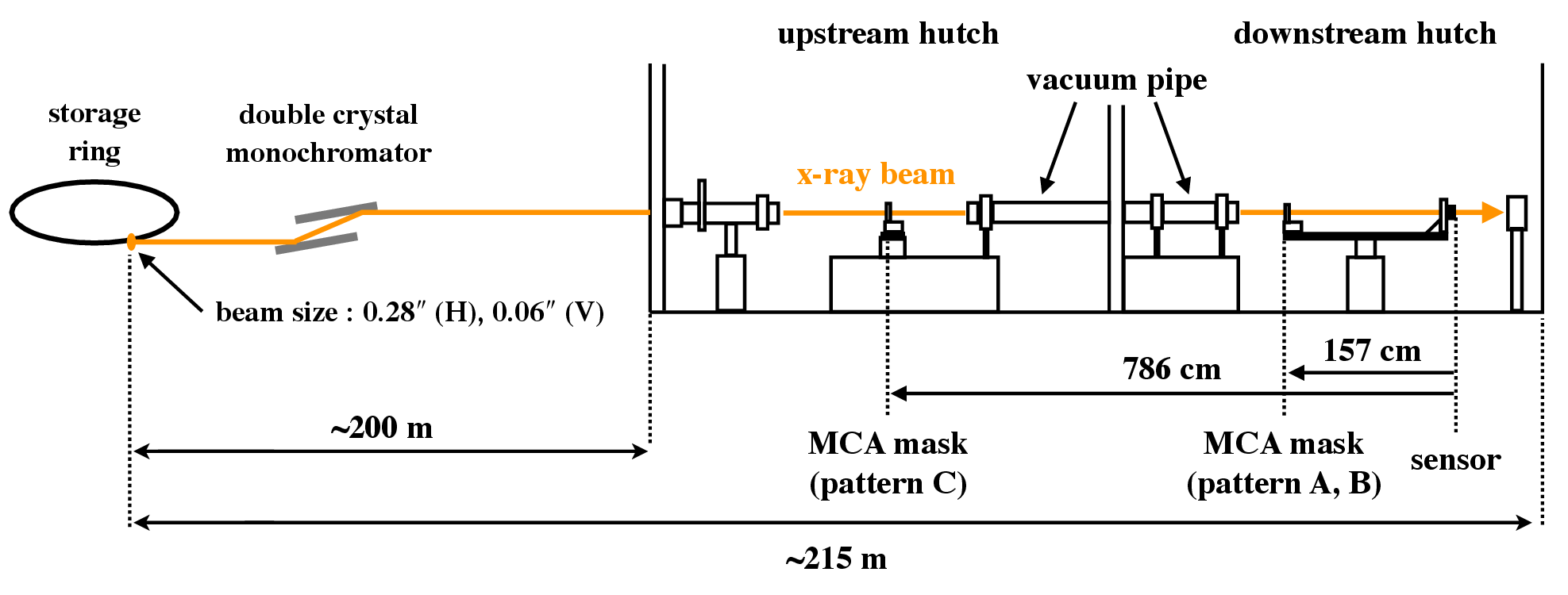}
   \end{tabular}
   \end{center}
   \caption[Overview of the entire experimental system.]  
   { \label{fig:SPring8_overview} 
   Overview of the entire experimental system. 
   BL20B2 provides the monochromatized synchrotron x-ray beam with a high degree of parallelization.
   Vacuum pipes are installed in the beam path to prevent air absorption.}
   \end{figure}

\begin{figure} [ht]
   \begin{center}
   \begin{tabular}{c} 
   \includegraphics[height=5.5cm]{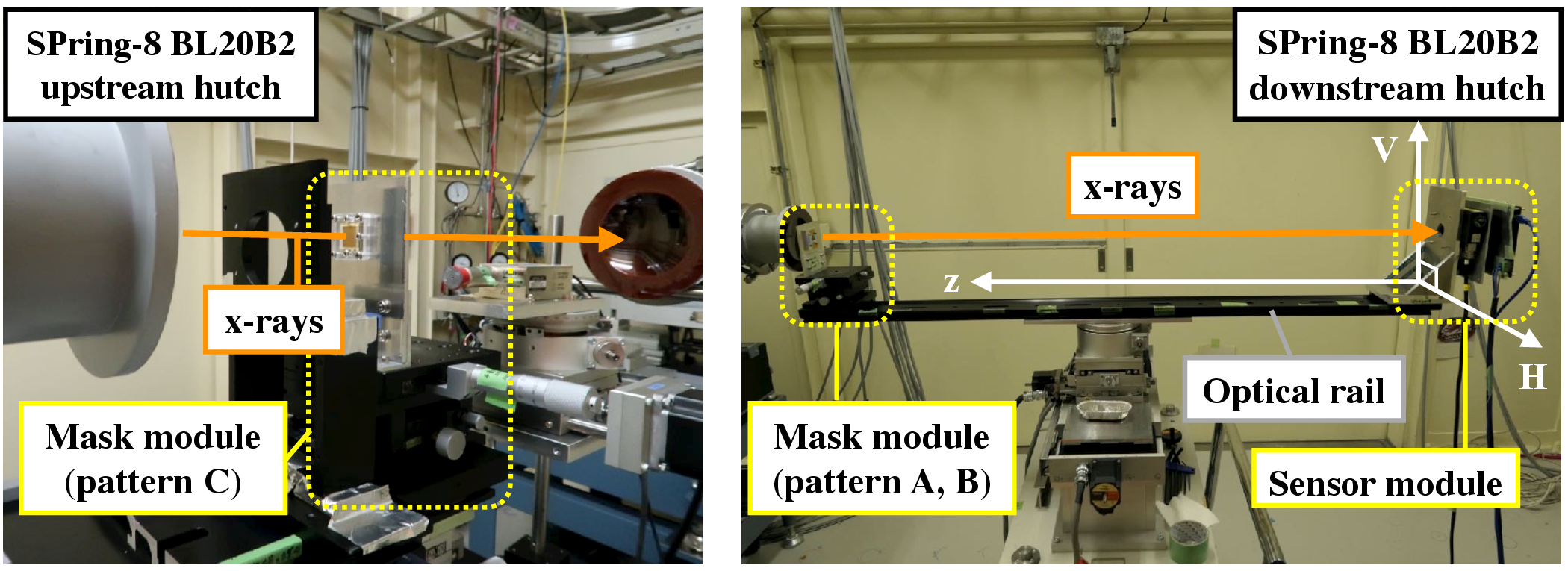}
   \end{tabular}
   \end{center}
   \caption[Photos of the experimental hutches.]  
   { \label{fig:SPring8_setup} 
   Photos of the upstream and downstream experimental hutches where we assembled the imaging system.
   White arrows indicate the coordinate system in the experimental hutches.
   Whereas the sensor module was fixed in the downstream hutch, 
   the mask module was installed in either the upstream or downstream hutches according to the distance $z_\mathrm{T}$.}
   \end{figure}

We conducted proof-of-concept experiments of high-resolution imaging with a MCA mask in SPring-8 BL20B2, a synchrotron radiation facility in Japan.
BL20B2 provides an x-ray beam with high intensity and a small divergence of $\ang{;;0.28}$ (H) and $\ang{;;0.06}$ (V).
Figure \ref{fig:SPring8_overview} illustrates a schematic overview of the entire experimental system, and Fig. \ref{fig:SPring8_setup} shows actual photos of the experimental hutches.
Throughout the experiments, the beam was monochromatized to $12.4\keV$ with a double crystal monochromator installed upstream of the experimental hutches,
and attenuation plates were inserted on the beam axis to adjust the beam intensity in order that the pile-up effect would be minimized to a negligible level.

A point to consider is that a fine-pitch MCA mask requires an x-ray detector with sufficiently high spatial and spectral resolutions
to resolve the self-images of aperture patterns and extract only x-rays with a wavelength of interest, respectively.
We employed the scientific CMOS sensor GMAX0505 in the new MIXIM as in the previous model of MIXIM.
GMAX0505 satisfies these requirements owing to the fact that its fine pixel size of $2.5\um$ is equivalent to the mask element sizes of the fine-pitch MCA masks
and that it has a high energy resolution of $176\eV$ (full-width at half maximum) at $5.9\keV$ at room temperature\cite{Asakura2019}.

As with the experiment with the previous model of MIXIM\cite{Asakura2023a}, 
a mask module and a sensor module were installed onto an optical rail so that the distance between them could be easily adjusted.
In the case of a spherical wave from a point source, Eq. \ref{eq:distance_plane} is modified as in the following formula:
\begin{linenomath}
\begin{align}
  \label{eq:distance_sphere}
  z_{\mathrm{T}} = m\frac{d^2}{\lambda} \frac{z_0}{z_0-\frac{md^2}{\lambda}} \; (m = 1, 2, 3...),
\end{align}
\end{linenomath}
where $z_0$ denotes the distance between the x-ray source and the MCA mask\cite{Patorski1988}.
We installed the mask module in a way that the configuration would satisfy Eq. \ref{eq:distance_sphere} ($m$=1) for 12.4-keV x-rays;
the mask-sensor distance $z_\mathrm{T}$ was set to $157\cm$ and $786\cm$ for the MCA masks with pitches of $12.5\um$ (patterns A and B) and $27.5\um$ (pattern C), respectively.

\subsection{Procedures}
\label{sec:experiment_3}

The analysis procedures with a MCA mask are basically the same as those with a multi-pinhole mask\cite{Asakura2023a};
we extracted x-ray events within the target energy range from the obtained frame data, 
created a photon count map on the sub-pixel coordinates of the x-ray events, and folded it with the best-estimated period
(we refer to it as a folded map; see Paper I for the sub-pixel analysis and folding-period estimation method).
In the case of the MCA mask, we need to decode the folded map to obtain an x-ray source profile.
An important point to note here is that the decoding process needs the actual unit aperture pattern of each MCA mask for $12.4\keV$ x-rays,
which differs from the originally-designed binary aperture patterns as shown in Fig. \ref{fig:micrographs}.
Hence, we also derived the transmittance map for each MCA mask by comparing the frame data in an experimental setup of a mask-detector distance of $1.58\cm$ 
(i.e., sufficiently close distance, with which diffraction effects are negligible) and those where the MCA mask was removed, and adopted it as the actual unit aperture pattern.

\begin{figure} [ht]
  \begin{center}
  \begin{tabular}{c} 
  \includegraphics[height=8cm]{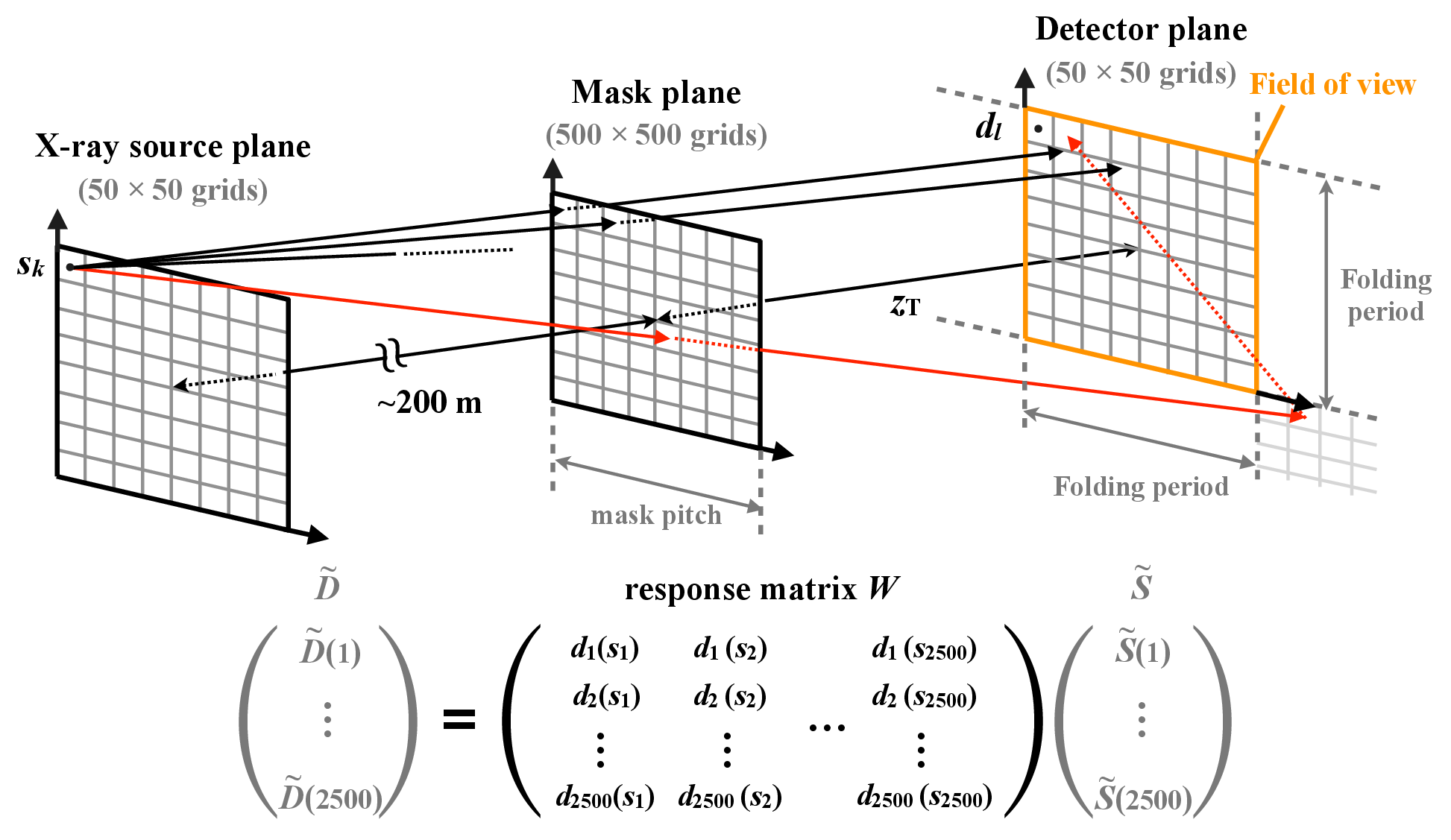}
  \end{tabular}
  \end{center}
  \caption[Overview of the geometry for deriving the decoding matrix.]  
  { \label{fig:ray-tracing} 
  Overview of the geometry for deriving the decoding matrix $W$. The red arrow represents the case where a photon arrives out of the field of view.}
\end{figure}

In this work, folded maps were decoded with the following procedure;
assuming that a background component is negligible, the response matrix $W$ follows
\begin{align}
  \label{eq:product}
  \left\{ \,
  \begin{array}{ll}
    \tilde{D} &= W \ast \tilde{S}, \\
    \tilde{D}(i) &= \sum\limits_{j} W(i, j) \tilde{S}(j),
  \end{array}
  \right.
\end{align}
where $\tilde{S}$ and $\tilde{D}$ are matrices which represent a source profile and a profile on a detector, respectively.
In our analysis, the response matrix $W$ was calculated under the assumption of a simple geometry (Fig. \ref{fig:ray-tracing});
a source plane and a detector plane were set to $50\times50$ grids, the same as a folded map, 
and a transmittance map, which was obtained from the experiment and was placed on a mask plane, was set to $500\times500$ grids.
Regarding a $k$-th source grid $s_k$ ($k = 1, 2, ..., 2500$) as a departure point, we calculated the trajectory of an x-ray photon passing through each mask grid, 
filled the transmittance at the mask grid into an $l$-th detector grid $d_l$ ($l = 1, 2, ..., 2500$) where the photon arrived, and then derived the average transmittance for each detector grid.
The position of an x-ray photon arriving out of the FOV such as the red arrow shown in Fig. \ref{fig:ray-tracing} was shifted 
by an integer multiple of the folding period to fit within the FOV, considering the folding procedure.
In this setup, the obtained pattern $d$ corresponded to the response vector for $s_k$.
The aforementioned calculation was repeated for all $s_k$, which yielded 2500 types of $d(s_k)$ in total.
The response matrix $W$ was finally obtained as a concatenation of each $d(s_k)$.
Fig. \ref{fig:ray-tracing} illustrates the configuration for our method.

Whereas classical coded aperture imaging often employs a correlation method to reconstruct an x-ray source profile,
such a simple method could cause blurring or artifacts especially in the case that the transmittance map does not have optimized patterns such as MURAs.
Hence, we adopted the Expectation-Maximization (EM) algorithm\cite{Dempster1977},
an algorithm for computing the maximum likelihood by repeating the Expectation-step (E-step) and Maximization-step (M-step) alternately.
In our case, E-step and M-step can be described as  
\begin{align}
  \label{eq:EM-algorithm}
  \left\{ \,
  \begin{array}{ll}
    \text{E-step: } \tilde{D}^{(n)}(i)    &= \displaystyle\sum\limits_{j} W(i, j) \tilde{S}^{(n)}(j) , \\
    \text{M-step: } \tilde{S}^{(n+1)}(j) &= \tilde{S}^{(n)}(j) \displaystyle\sum\limits_{i} \dfrac{\delta(i) W(i, j)}{\tilde{D}^{(n)}(i)},
  \end{array}
  \right.
\end{align}
where $n$ and $\delta$ represent the iteration number and obtained folded map, respectively\cite{Shepp1982}.
Given that the likelihood never decreases at each iteration (see Appendix of Ref. \citenum{Shepp1982}), 
the likelihood converges to the maximum after a repetition of the E-step and M-step;
accordingly, we can estimate the most probable x-ray source profile with this algorithm.
In this work, we employed, as the measure of convergence, the Kullback-Leibler (KL) divergence\cite{Kullback1951}, defined by the following equation:
\begin{align}
  \label{eq:KL-divergence}
  KL(D^{(n+1)}, D^{(n)}) = \sum\limits_{i} D^{(n+1)}(i) \log \frac{D^{(n+1)}(i)}{D^{(n)}(i)},
\end{align}
and repeated these steps until the KL divergence dropped below $10^{-10}$.
As the initial condition of this iterative algorithm, we employed a spatially-uniform source profile in our analysis.

\section{Results}
\label{sec:results}  

\subsection{Reconstructed Source Profiles}
\label{sec:results_1}

\begin{figure} [ht]
   \begin{center}
   \begin{tabular}{c} 
   \includegraphics[height=5.5cm]{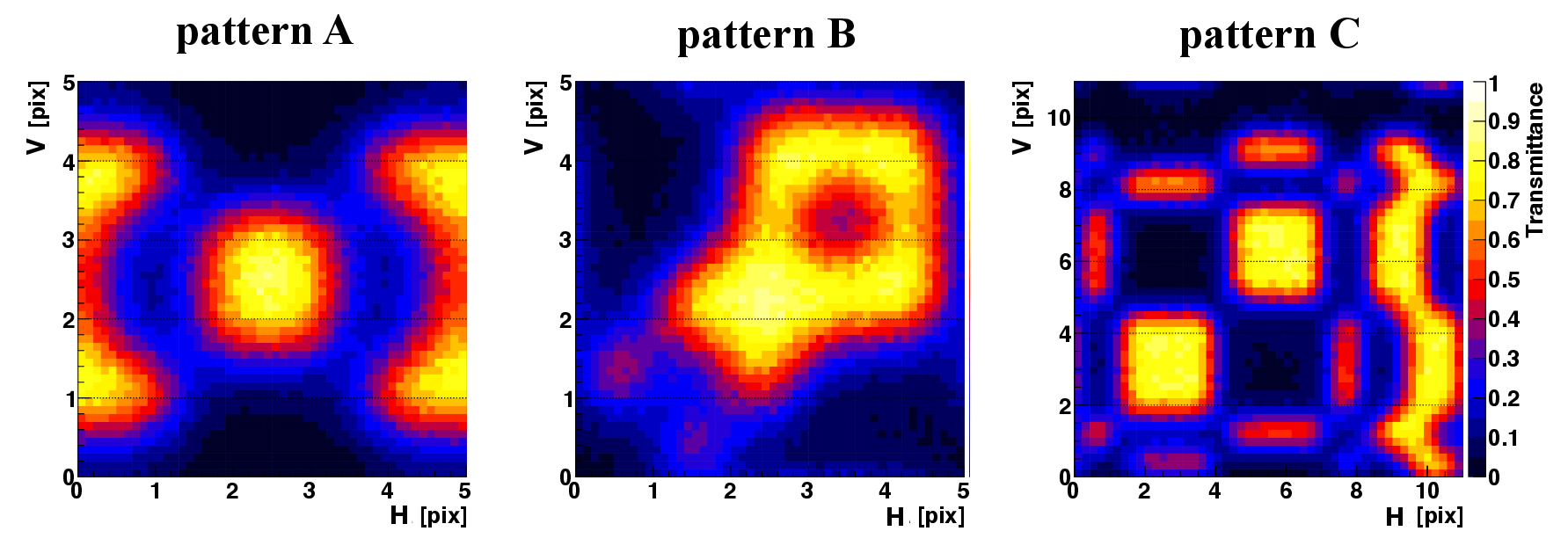}
   \end{tabular}
   \end{center}
   \caption[Transmittance maps for patterns A to C at $12.4\keV$.]  
   { \label{fig:TransmittanceMap} 
   Transmittance maps for patterns A to C from the left to right panels, respectively, at $12.4\keV$.
   The horizontal and vertical axes show the best-estimated folding periods in units of the sensor pixel size.
   All of them have the overall transmittance of over $30\%$.}   
   \end{figure}

Figure \ref{fig:TransmittanceMap} shows the transmittance map of each MCA mask at $12.4\keV$ (binned with $50\times50$ pixels).
They are slightly different from those of the micrographs shown in Fig. \ref{fig:micrographs},
but their sharpness implies that the masks certainly have equally-spaced structure, which is essential for the Talbot effect.
Patterns A, B, and C have the overall transmittance of $31.1\%$, $33.9\%$, and $30.0\%$, respectively. 
These values are much higher (over 20 folds) than that with the previous model of MIXIM with a multi-pinhole mask ($\sim1.3\%$).
Ideally, they should agree with the opening fraction of the original designs ($\sim50\%$).
In reality, the values are lower than that presumably because of imperfection of fabrication, and so the values can be improved further in the future.

\begin{figure} [ht]
   \begin{center}
   \begin{tabular}{c} 
   \includegraphics[height=10cm]{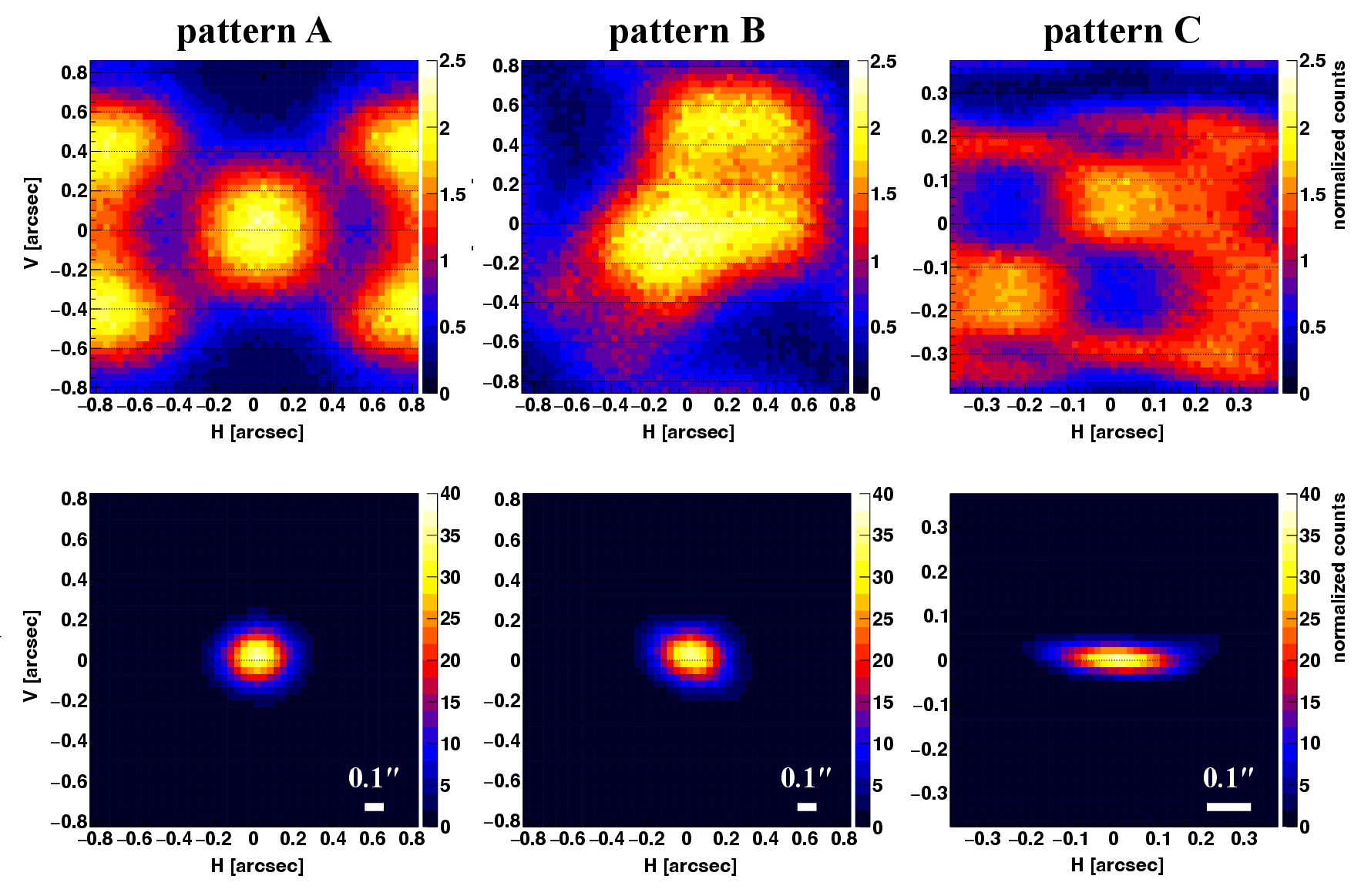}
   \end{tabular}
   \end{center}
   \caption[Folded maps and decoded maps for patterns A to C at $12.4\keV$.]  
   { \label{fig:SingleSrc_result}
   (Top) Folded maps and (bottom) decoded maps obtained with the MCA masks at $12.4\keV$.
   The horizontal and vertical axes denote the incident angle, and the color scale shows the normalized counts.
   Note that the FOV for pattern C is different from the others due to a longer mask-detector distance.}
   \end{figure}

The x-ray beam in our detection system can be regarded as a point source, given that its spatial extent is comparable to the theoretical angular resolution.
Hence, the folded maps at $m=1$ should have the same profiles as these transmittance maps for patterns A and B. 
The obtained folded maps at $z=157\cm$ (patterns A and B) and $z=786\cm$ (pattern C) are shown in upper panels of Fig. \ref{fig:SingleSrc_result} 
(each of them contains about one million photons in total),
which explicitly demonstrate that even masks with complex aperture patterns can form self-images for a point-source in the x-ray band.

Applying the decoding procedure described in section \ref{sec:experiment_3} to the folded maps, 
we reconstructed x-ray source profiles (hereafter, referred to as decoded maps), as shown in the lower panels of Fig. \ref{fig:SingleSrc_result}.
They were binned in $50\times50$ pixels, normalized by the average counts per pixel, and additionally smoothed with a Gaussian filter ($\sigma \sim 1$ pixel) for better visualization.
The MIXIM configuration with patterns A and B have a FOV of $\ang{;;1.66}\times\ang{;;1.66}$ while that with pattern C has a FOV of $\ang{;;0.75}\times\ang{;;0.75}$,
and all of them successfully capture the image of the x-ray beam, with negligibly small sidelobes (less than 1\% of the peak height).
Notably, whereas the decoded maps of patterns A and B converge to a point-like source, that of pattern C shows a horizontally-elongated profile,
which is in reasonably good agreement with the beam divergence mentioned in section \ref{sec:experiment_2}.
This fact indicates that pattern C achieved a much higher angular resolution than the others because of a long mask-detector distance of $786\cm$.

\subsection{Separation of Two Point-Sources}
\label{sec:results_2}

\begin{figure} [ht]
   \begin{center}
   \begin{tabular}{c} 
   \includegraphics[height=5cm]{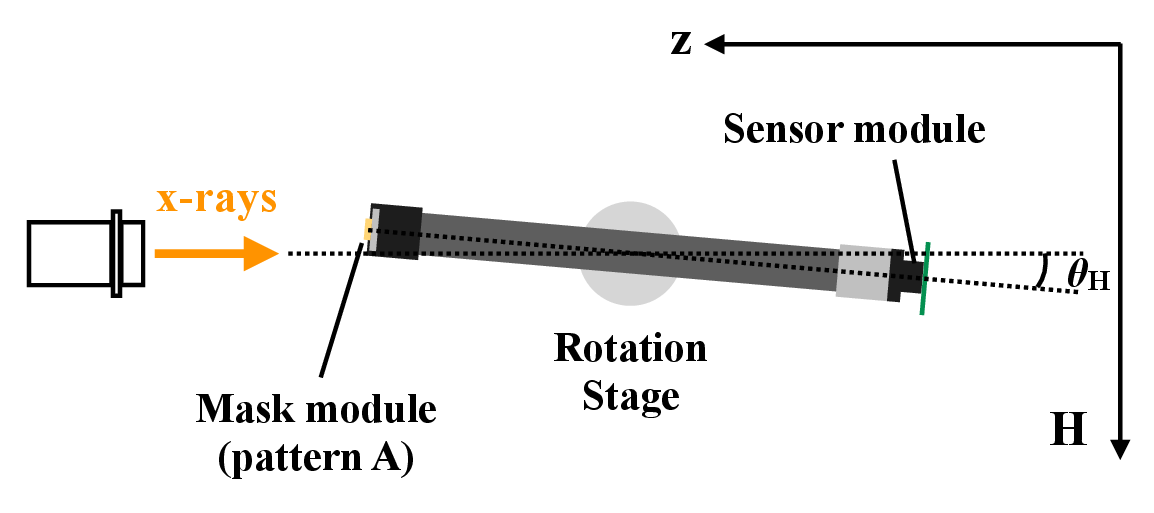}
   \end{tabular}
   \end{center}
   \caption[Overview of the configuration for the two point-source observation.]  
   { \label{fig:TwoPointSrc_config} 
   Overview of the experimental configuration for the two point-source observation.
   The optical rail was rotated in the horizontal direction in steps of $\ang{;;0.108}$ for data acquisition.}
   \end{figure}

\begin{figure} [ht]
   \begin{center}
   \begin{tabular}{c} 
   \includegraphics[height=10cm]{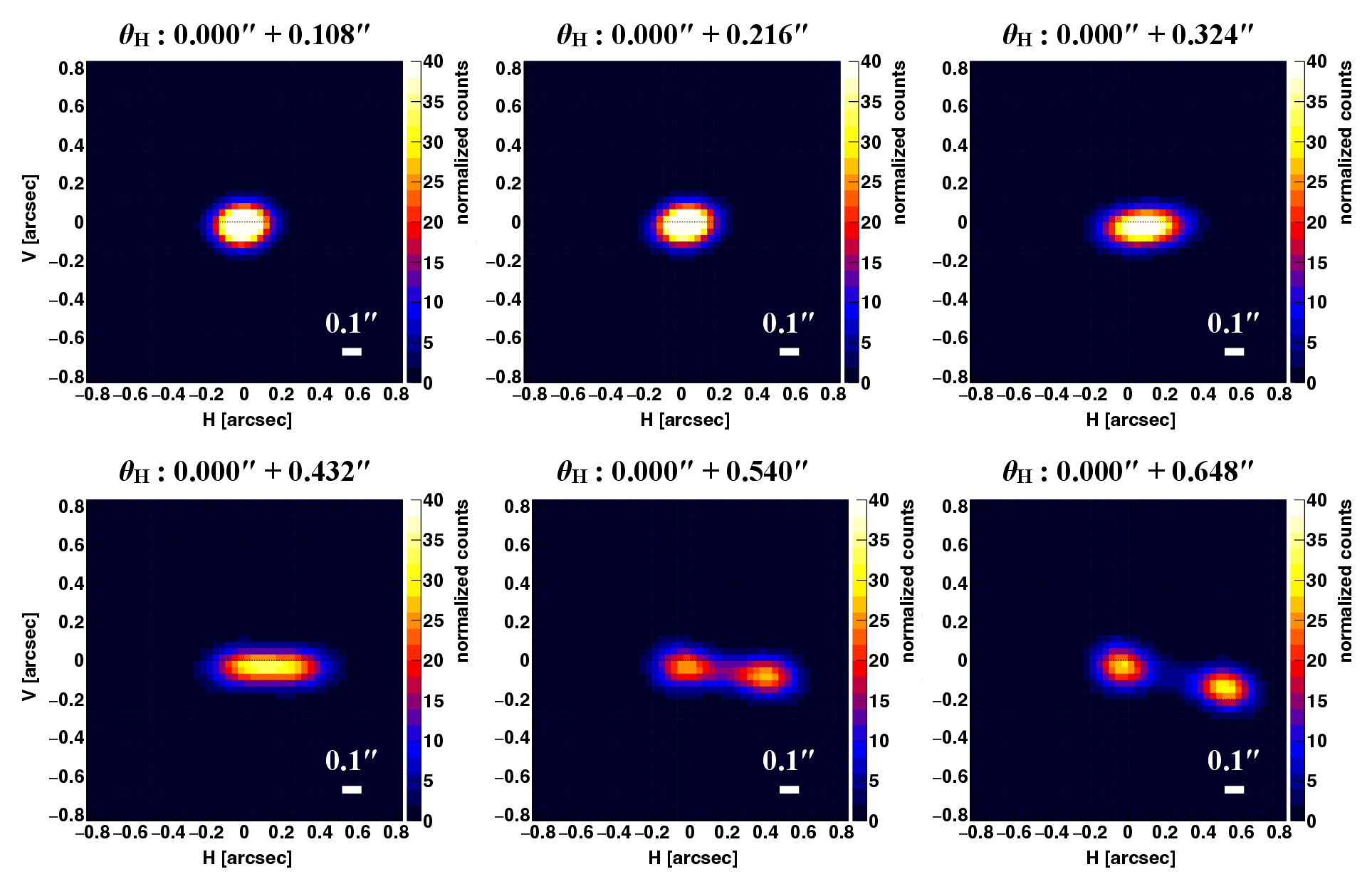}
   \end{tabular}
   \end{center}
   \caption[Decoded maps assuming the two point-source observation.]  
   { \label{fig:TwoPointSrc_result} 
   Decoded maps derived from both the on-axis and off-axis data, 
   assuming the observation of identical two point-sources apart by $\theta_{\mathrm{H}}$.}
   \end{figure}

The extent of the source profiles obtained in the previous subsection should roughly represent the angular resolution of the employed MCA mask.
However, given that the result was obtained by means of a complicated reproduction procedure, 
it is not trivial how multiple targets or a spatially extended target are observed with MIXIM with a MCA mask.
We here evaluated the angular resolution of multiple point-sources with pattern A by simulating an observation of two sources as follows.
First, we rotated the optical rail in the horizontal direction by angles $\theta_{\mathrm{H}}$ in steps of $\ang{;;0.108}$ as illustrated in Fig. \ref{fig:TwoPointSrc_config}
 and obtained frame data at each step (n.b., the number of frames and exposure were the same for each step).
Second, we created a folded map with x-ray events extracted from two datasets: the on-axis ($\theta_{\mathrm{H}} = 0$) and off-axis ($\theta_{\mathrm{H}} \neq 0$) data.
Then, decoding this folded map practically simulated the observation of two point-sources apart by $\theta_{\mathrm{H}}$.
Figure \ref{fig:TwoPointSrc_result} shows the decoded maps for $\theta_{\mathrm{H}}$ varied from $\ang{;;0.108}$ to $\ang{;;0.648}$.
The two sources are correctly reconstructed for each decoded map in terms of both the spatial structure and brightness;
the image is apparently elongated unlike a point source in the reconstructed map for $\theta_{\mathrm{H}} \geq \ang{;;0.324}$,
and the image profiles of the two sources are clearly separated for $\theta_{\mathrm{H}} \geq \ang{;;0.540}$.
These results support that the MCA mask realizes the angular resolution close to the theoretical expectation.

\subsection{Energy Dependence}
\label{sec:results_3}

\begin{figure} [ht]
   \begin{center}
   \begin{tabular}{c} 
   \includegraphics[height=9.5cm]{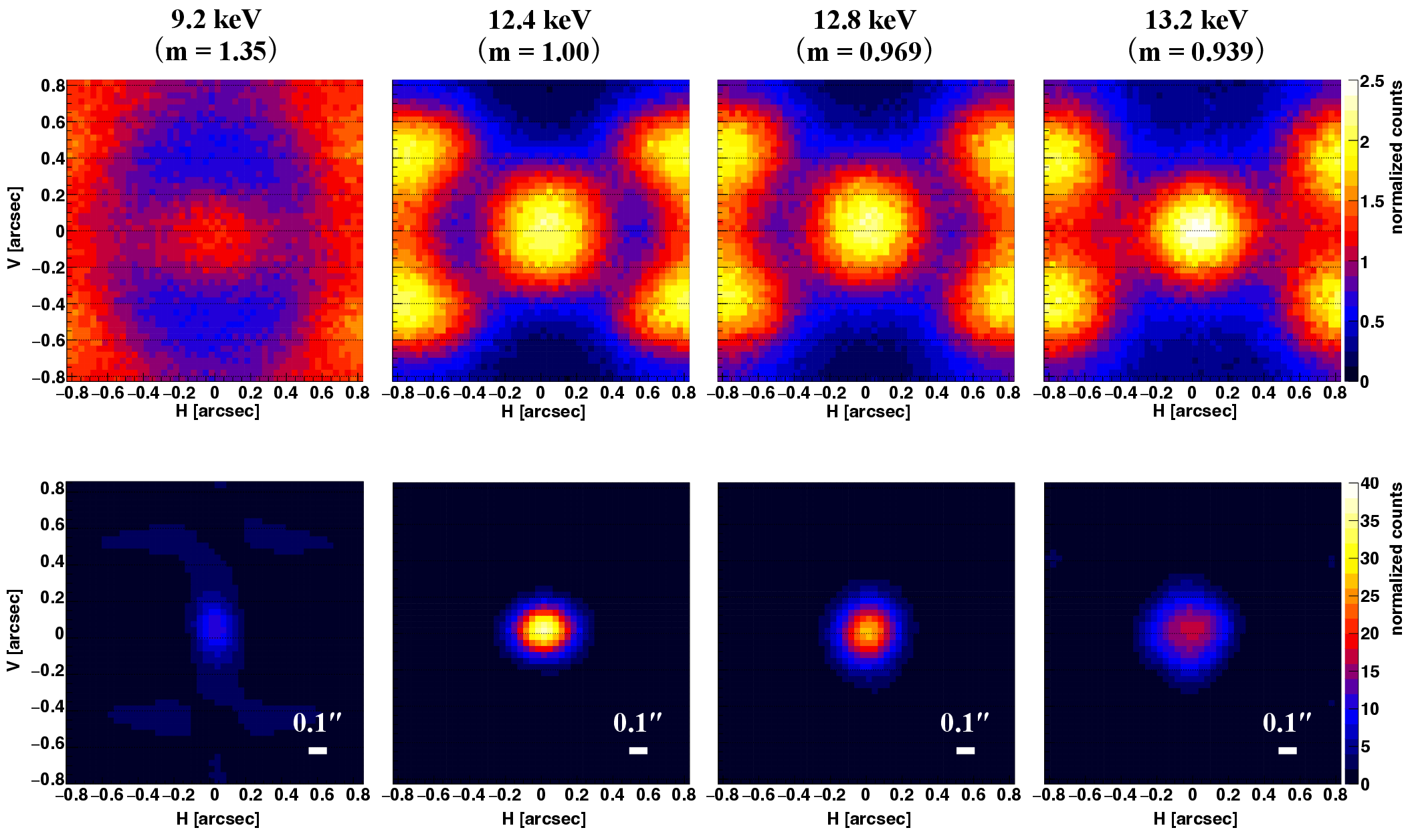}
   \end{tabular}
   \end{center}
   \caption[Folded maps with a variety of beam energies.]  
   { \label{fig:E_dependence} 
    (Top) From the left panel: folded maps with a beam energy of $9.2\keV$ ($m=1.35$), $12.4\keV$ ($m=1.00$), $12.8\keV$ ($m=0.969$), and $13.2\keV$ ($m=0.939$).
    (Bottom) decoded maps derived with the folded maps in the top panels and the transmittance map in Fig. \ref{fig:TransmittanceMap}.}
   \end{figure}

When we observe a non-monochromatic x-ray source, x-rays out of the target wavelength deteriorate the quality of self-imaging.
As $m$ is offset from integers, self-images with a multi-pinhole mask are simply blurred, 
but those with MCA masks are supposed to be complexly varied according to interference.
This fact implies that an energy shift might cause not only the deterioration of angular resolution but also the wrong reproduction of a source profile.
To investigate the effect of non-monochromatic source x-rays on folded and decoded maps 
(n.b., all of them were deciphered with the transmittance map shown in Fig. \ref{fig:TransmittanceMap}),
we made experiments with various x-ray beam energy shifted from $12.4\keV$ for pattern A with the same configuration.
The top and bottom panels of Fig. \ref{fig:E_dependence} show the obtained folded and decoded maps at a variety of beam energies.
Notably, the decoded maps maintain a point-like profile without substantial side-lobes 
at least within an energy range of $\pm5\%$, whereas they are gradually blurred as $m$ deviates from unity.
Meanwhile, the profile at $9.2\keV$ has a much lower visibility than the others,
which demonstrates that the Talbot effect certainly plays an important role to obtain clear self-images.

\section{Discussion}
\label{sec:discussion}  

\subsection{Angular resolutions of the MCA masks}
\label{sec:discussion_1}

The angular resolution of a multi-pinhole system as in the previous MIXIM can be simply calculated from its point spread function (PSF).
However, the same procedure cannot be applied to the MCA configuration 
because PSF-deconvoluted x-ray beam images are directly obtained from the observed images without the use of a PSF or similar.
This fact implies that the angular resolution is not equal to the simple divergence and must be somehow estimated without PSF information.
In this experiment, however, the images obtained with patterns A and B are wider in the vertical direction than the x-ray beam divergence,
the primary reason of which is presumably the spatial resolution of the sensor is insufficient to resolve the broadening of the self-images in the case of $z=157\cm$.
In this case, a simple Gaussian fit gives their vertical angular widths to be $\ang{;;0.2}$ (full-width at half-maximum), which can be interpreted as the angular resolution.
We conclude that our novel prototype with a compact system size of $\sim1.5\m$ achieved an angular resolution of $\ang{;;0.2}$.
This angular resolution is better than the expected performance described in section \ref{sec:MCA};
we consider that this is because both the transmittance maps and folded maps have patterns finer than the designed mask-element size.

Notably, a finer aperture pattern can be fabricated through x-ray lithography with an electron beam instead of currently-used lithography with a laser.
A finer aperture pattern would enable us to further improve an angular resolution, providing that it is combined with an x-ray detector with a sufficiently high spatial resolution 
and that the number of the mask elements (i.e., a pitch of a mask) is accordingly increased to retain the mask-sensor distance.
The $53\times53$ MURA pattern adopted for \textit{INTEGRAL}\cite{Goldwurm2003} is a real example of the increased number of the mask elements.

\subsection{Imaging Performance with Poor Photon Statistics}
\label{sec:discussion_2}

The x-ray beam at BL20B2 provided sufficient photon statistics even in a short observation time.
In most actual observations of astrophysical x-ray sources, however, photon statistics are greatly limited,
and therefore, it is necessary to decipher folded maps as efficiently as possible to derive meaningful results.
To evaluate the observation feasibility with limited photon statistics, we re-analyze the x-ray events used in section \ref{sec:results_2}, varying the total x-ray event number $N$ as follows.
We randomly extract $N/2$ x-ray events from both the on-axis ($\theta_{\mathrm{H}} = \ang{;;0.00}$) and off-axis ($\theta_{\mathrm{H}} = \ang{;;0.54}$) data, 
create a folded map with these events, and decipher it with the same decoding process as applied in section \ref{sec:results_2}.
Repeating this process 100 times yields 100 independent decoded maps.
Finally, we adopt the mean and standard deviation of the results as the reconstructed source profile and its standard error.

\begin{figure} [ht]
   \begin{center}
   \begin{tabular}{c} 
   \includegraphics[height=9cm]{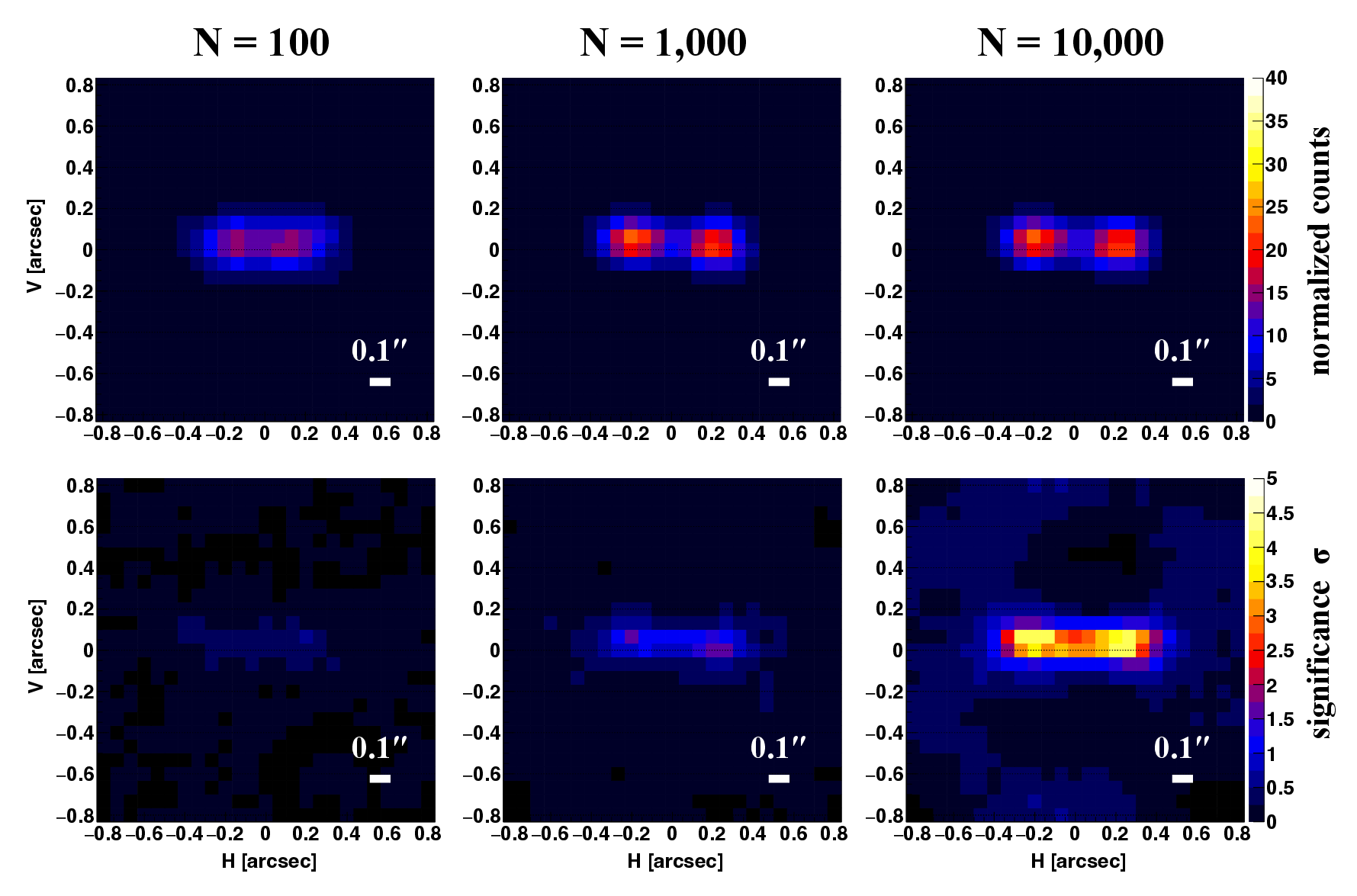}
   \end{tabular}
   \end{center}
   \caption[Decoded maps obtained with limited photon statistics.]  
   { \label{fig:PoorPhotonStats} 
   (Top) Decoded maps derived in the same way as the case of $\theta_{\mathrm{H}}=\ang{;;0.54}$ in Fig. \ref{fig:TwoPointSrc_result}, 
   but the total x-ray counts $N$ (displayed at the top) are reduced for the aim of simulating more realistic space-observation situations with expected limited photon statistics. 
   The color levels are normalized by the average counts per pixel. 
   (Bottom) Same as the top panels, but the color scale is normalized by the standard error per pixel to show a significance level.}
   \end{figure}

Figure \ref{fig:PoorPhotonStats} shows the decoded maps with $N$ of $10^2$, $10^3$, and $10^4$ (binned with $25\times25$ pixels).
Whereas the two sources with $N$ of $10^4$ are clearly reconstructed with a separation of $\ang{;;0.5}$, those with $N$ of $10^2$ have relatively blurred profiles,
the difference of which is qualitatively expected given that the spatial variation of the images increases as $N$ decreases.
This result implies that poor photon statistics result in an increase in the uncertainty of source positions or their profiles,
whereas the pixels corresponding to the source positions have a confidence level of more than $3\sigma$ for $N$ of $10^4$.
We also evaluated the tolerance to background photons by adding some photons to randomly-selected pixels of the folded maps before the decoding process,
and confirmed that the reconstructed source profiles have almost same as Fig. \ref{fig:PoorPhotonStats} when the number of background photons is at least less than 10\% of N.

\subsection{Future Prospects for Performance Improvement}
\label{sec:discussion_3}

A finer aperture pattern (and a sensor with a smaller pixel size) would improve an angular resolution, as mentioned in section \ref{sec:discussion_1}.
In reality, the decoding process, in addition to the mask and sensor parameters, also significantly affects the imaging performance, especially when the target x-ray sources have a complicated spatial profile.
In this work, we assumed a spatially-flat pattern as the initial source profile in our decoding algorithm.
Ideally, we should more or less know the spatial profile, potentially beyond the FOV, of an x-ray source in advance and employ it for accurate reconstruction of the image.
It is, however, difficult for MIXIM with a MCA mask to obtain the large-scale structure of targets to observe because it has a narrow FOV.
For this reason, MIXIM in actual deployment in space should be equipped with several MCA masks with different pitches to realize the different FOVs observed at the same time, 
and such MCA masks have mutually different mask-sensor distances to maintain interferometry for a target wavelength.
We note that the target structure with a scale of $\ang{;;0.5}$ in the soft x-ray band can be obtained with \textit{Chandra}, 
the information of which would be also useful for the assumed initial profile for image reconstruction with MIXIM.
The imaging performance also depends on artifacts ascribed to aperture patterns; they would be reduced if each mask has a different pattern, 
as demonstrated in Ref. \citenum{Kasuga2020}, which is another advantage to adopting different patterns to the masks.
Furthermore, the standard EM-algorithm that we adopted in this work has room for improvement with regard to, e.g., the convergence speed and error estimates\cite{Isaac1989}.
Optimization of the aperture patterns and decoding process is desirable for actual observations in the future to obtain the best results.

\section{Summary}
\label{sec:summary}  

MIXIM, our novel x-ray imaging system, in the previous configuration with a multi-pinhole mask and CMOS sensor 
achieved a high angular resolution of $\ang{;;0.5}$ with a system size of as small as $\sim1\m$ in the past proof-of-concept experiments.
However, its small effective area due to a very low opening fraction is a serious problem, especially for x-ray astronomical observations with low photon fluxes.
To address the problem and improve the effective area, we newly employed MCA masks (equally-spaced coded aperture patterns with an opening fraction of $\sim50\%$) to MIXIM
and conducted proof-of-concept experiments at SPring-8 BL20B2 to evaluate their imaging performance.
These experiments demonstrated that our prototype successfully obtained the encoded patterns of the x-ray source profile,
which indicates that the Talbot effect works for the MCA masks even though their aperture patterns are complex.
Then we successfully reconstructed the x-ray source profile, deciphering the encoded patterns with our decoding algorithm with the original aperture patterns.
The successful reconstruction implies that a system with a MCA mask (with a pitch of $12.5\um$) realizes both a high angular resolution of $\ang{;;0.2}$ and a compact system size of $\sim1.5\m$.
Thus, the high opening fraction of a MCA mask achieves $\sim25$ times as large an effective area in MIXIM as that with a multi-pinhole mask.
Given the severe technical constraint in practice on the physical size of any x-ray imaging system to be deployed in orbit and the requirements for them to be sensitive enough for faint targets,
the introduction of a MCA mask to MIXIM considerably increases its feasibility for future deployment for astronomical observations.

\subsection*{Disclosures}
The authors have no conflicts of interest to disclose.

\subsection*{Data Availability}
The data that support the findings of this study are available from the corresponding author upon reasonable request.

\subsection* {Acknowledgments}

The authors are grateful to S. Sakuma, A. Ishikura, K. Okazaki, M.Hanaoka, K. Hattori, K. Sawagami, T. Mineta, Y. Matsushita, W. Kamogawa, Y. Ode (Osaka University), 
Dr. M. Hoshino, and Dr. K. Uesugi (JASRI) for their support with the proof-of-concept experiments.
The first author is also grateful to his current affiliation, Japan Aerospace Exploration Agency (JAXA).
This work was supported by Japan Society for the Promotion of Science (JSPS) KAKENHI Grant Nos. 20J20685 (KA), 18K18767, 19H00696, 19H01908, 20H00176 (KH), 21K20372 (TY), 20H00175, 23H00128 (HM).
The synchrotron radiation experiments were performed at the BL20B2 of SPring-8 with the approval of the Japan Synchrotron Radiation Research Institute (JASRI) 
with Proposal Nos. 2018A1368, 2018B1235, 2019A1503, 2019B1492, 2020A1506, 2021A1442 (KH) and 2022A1256 (HM).
Part of the results of this manuscript was reported in SPIE proceedings\cite{Asakura2022}.


\bibliography{report}   
\bibliographystyle{spiejour}   

\vspace{2ex}\noindent\textbf{Kazunori Asakura} is a researcher in Japan Aerospace Exploration Agency. 
He received his BS, MS and Ph.D degrees from Osaka University in 2018, 2020 and 2023, respectively. 
His current research interests include x-ray astronomy and instrumentation for x-ray imaging spectroscopy and polarimetry.

\vspace{1ex}
\noindent Biographies and photographs of the other authors are not available.

\newpage
\listoffigures

\end{spacing}
\end{document}